\documentclass{article}
\usepackage{arxiv}

\usepackage[utf8]{inputenc} 
\usepackage[T1]{fontenc}    
\usepackage{hyperref}       
\usepackage{url}            
\usepackage{booktabs}       
\usepackage{amsfonts}       
\usepackage{nicefrac}       
\usepackage{microtype}      
\usepackage{lipsum}
\usepackage{graphicx}
\graphicspath{ {./images/} }
\usepackage{natbib}
\usepackage{float}
\usepackage{adjustbox}

\usepackage{xcolor}
\hypersetup{
    colorlinks=true,
    citecolor=teal,
    linkcolor=teal,
    filecolor=teal,
    urlcolor=teal,
}

\usepackage{xspace}
\definecolor{trainingcolor}{RGB}{31,119,180}
\definecolor{testcolor}{RGB}{255,127,14}
\definecolor{deferredcolor}{RGB}{44,160,44}

\newcommand{\GalaxyZoo}{{\tt GalaxyZoo}\xspace}
\newcommand{\HateSpeech}{{\tt HateSpeech}\xspace}
\newcommand{\CifarH}{{\tt Cifar10H}\xspace}

\newcommand{\ChestXRay}{{\tt ChestXRay}\xspace}

\usepackage[capitalize, nameinlink]{cleveref}

\title{Too Much of the Same:\\From Algorithmic to Human Bias in Learning to Defer}
\author{
 Dario Pesenti \\
  CIMeC, University of Trento\\
  \texttt{dario.pesenti@unitn.it} \\
   \And
 Alessandro Bogani \\
  DISI, University of Trento\\
  \texttt{alessandro.bogani@unitn.it} \\
  \AND
  Stefano Teso \\
  CIMeC \& DISI, University of Trento\\
  \texttt{stefano.teso@unitn.it} \\
  \And
  Andrea Pugnana \\
   DISI, University of Trento\\
  \texttt{andrea.pugnana@unitn.it} \\
}

\begin{document}
\maketitle

\begin{abstract}
Learning to Defer (LtD) extends supervised learning by allowing a Machine Learning (ML) model to defer harder or less confident decisions to a human expert. 
Despite being geared for human-AI collaboration, LtD strategies neglect the potential negative interference of human cognitive biases.
Our contribution is twofold.
First, we demonstrate that standard LtD strategies show class-dependent sampling bias in classification tasks in practice, and thus may disproportionately defer the minority classes when applied to imbalanced datasets.
Second, we show that such asymmetries in task delegation may trigger human biases, ultimately leading to poorer downstream decision making.
Specifically, we conduct a user study ($N=226$) where participants complete a classification task on a set of deferred items, with conditions presenting different levels of class imbalance.  Our results show that participants exposed to a highly imbalanced rejection set achieved lower classification accuracy in the majority class compared to those exposed to a more balanced set, regardless of which class constituted the majority.
Exploratory analyses suggest that this may be an instance of the Test-taker's effect, which stems from a mismatch between the actual distribution of classes and the participants' expectations about that distribution.
Finally, we discuss the implications of these findings for the deployment of LtD algorithms.
\end{abstract}

\section{Introduction}

Various real-world applications characterized by potentially high levels of cognitive fatigue, from dataset annotation to healthcare, have seen an increasing use of ML models to aid decision-making \citep{soori2024ai}.  Since a complete delegation of decisions to AI systems is often undesirable due to ethical and technical concerns \citep{act2024eu}, it is preferable to deploy some form of \textit{hybrid decision making} (HDM) for reducing cognitive load while retaining human oversight \citep{mansoul2013hybrid}.

One of the most widely studied approaches to HDM is \textit{learning to defer} (LtD)~\citep{madras2018predict}. In LtD, the ML model evaluates its ability to handle a given item correctly, and defers those that are too difficult or risky to handle in autonomy to a human expert.
The decision to defer can be based on predetermined metrics (e.g., the novelty of the input or the confidence of the predictor), and evaluated against a threshold \citep{van2021reject, hendrickx2024machine}; alternatively, a separately trained ``rejector'' model can be employed
to route decisions to another agent \citep{mozannar2020consistent}; see \citet{punzi26} for an extensive review.

By allowing experts to focus on items where their judgment is most required, LtD algorithms can reduce cognitive load while simultaneously increasing human-AI team performance compared to either the human or the model in isolation \citep{hendrickx2024machine, hemmer2023human}.
For this reason, they have found application in, e.g., biomedicine \citep{kompa2021second} and vehicle design testing \citep{hendrickx2022know}.

Here, we focus on a critical but neglected issue: \textit{algorithmic biases implicit in the LtD design can trigger human cognitive biases, impacting down-stream decision making}.
A first issue is that, when applied to imbalanced decision tasks, LtD strategies tend to disproportionately defer the more challenging minority class(es) over the others~\citep{DBLP:conf/aistats/PugnanaR23,DBLP:journals/dmlr/PugnanaPDR24}.
A second issue is that
this entails that the human decision maker will be faced with an abundance of same-class items.
This can be problematic in two ways:
\textit{i}) It may produce attentive and decision-making biases, \citep{wolfe2013prevalence}, associated with prolonged monitoring \citep{head2014sustained}, and skewed target prevalence \citep{wolfe2013prevalence}, know as \textit{Prevalence effect}; this may reduce performance on \textit{minority}-class instances due to users' limited exposure to them. E.g., in airport screening, repeated exposure to innocuous objects may increase the likelihood that a dangerous object is incorrectly classified as harmless.
\textit{ii}) It may generate uncertainty due to the discrepancy between users' expectation about the distribution of deferred items and the distribution they actually encounter. This is known as \textit{Test-taker's effect} \citep{lee2019test}, owing to the fact that students faced with a multiple-choice exam with simple answer patterns (e.g., the correct answer repeatedly appearing in the same position) violate their expectations, leading them to revise their initial judgments and thereby reduce their accuracy \citep{paul2014}.

We study empirically both issues.
On the computational side, we tested three LtD strategies -- Realizable Surrogate \citep{mozannar2023should}, Selective Prediction \citep{geifman2017selective}, and Compare Confidence \citep{raghu2019algorithmic} -- on one naturally imbalanced dataset \ChestXRay \citep{wang2017chestx} and three other standard LtD datasets \GalaxyZoo, \HateSpeech, and \CifarH where we manipulated the class composition in the training set, showing that, when this is sufficiently imbalanced, rejected items can comprise up to $100\%$ of the instances from the minority class.
To test whether this form of sampling bias affects decision makers, we carried out a user study ($N=226$ participants, $33'900$ total observations) emulating an LtD scenario, in which one specific class gets overly deferred at a time, and compare it to a control condition with a more balanced set of deferred items.  Our findings suggest that participants exposed to a heavily imbalanced rejection set are overall less accurate than participants exposed to a more balanced set.
Follow-up exploratory analyses favor the Test-taker's effect over the Prevalence effect as an explanation for our findings, in that users' performance drops on the rejection set's \textit{majority} class, which our data suggest to be due to a discrepancy between the expected and actual distribution of the items to be classified.
Overall, our work challenges the common LtD assumption that human performance is independent of the deferral policy: in imbalanced settings, LtD may alter human accuracy and risk reducing overall human–AI team performance by skewing the distribution of deferred cases.

\paragraph{Contributions.}  Summarizing:
\begin{itemize}
    \item We highlight a potential issue with LtD, consisting in biased deferral sampling towards specific classes, especially if trained on imbalanced datasets. 
    \item We verify empirically that state-of-the-art LtD strategies are prone to producing imbalanced rejection sets, thus potentially triggering cognitive effects.
    \item We conduct a user study showing that imbalanced rejection sets can reduce performance on the majority class, with potential downstream consequences.
\end{itemize}

\section{Preliminaries \& Related Work}
\label{sec:preliminaries}

\paragraph{Learning to defer.}

LtD~\citep{madras2018predict} extends supervised learning techniques by allowing a ML model to \textit{defer} its prediction to another predictor, e.g., a human expert or a better ML model~\citep{DBLP:conf/aaai/RuggieriP25}.

Initial works proposed heuristics to achieve such a goal~\citep{raghu2019algorithmic,wilder2021learning}. More recently, several works have studied how to learn deferring systems with theoretical guarantees, including~\citep{mozannar2020consistent,okati2021differentiable,verma2022calibrated,DBLP:conf/icml/CharusaieMSS22,mozannar2023should,DBLP:conf/nips/CaoM0W023,DBLP:conf/aistats/LiuCZF024,DBLP:conf/aaai/GaoY25,DBLP:conf/aaai/LiLWMWB26}.

Recent directions have extended the original LtD formulation studying multiple-experts~\citep{DBLP:conf/aistats/VermaBN23,mao2023two,DBLP:journals/corr/abs-2504-12988,DBLP:conf/icml/MaoM025,DBLP:conf/aaai/ZhangNWDRC26,Liu26}; multi-task settings~\citep{Pugnana25,DBLP:conf/icml/MontreuilHCNO25}; robustness~\citep{DBLP:conf/icml/MontreuilCNO25,DBLP:journals/tmlr/FangN26}; rejection of inexplainable decisions \citep{stradiotti2025learning}; and applications~\citep{DBLP:conf/aaai/StrongMN25,montreuiloptimal}.

\paragraph{Terminology}
In the following, we refer to the set of items used for training the model as the \textit{training set}, while the set of all of the deferred test items is the \textit{rejection set}. The number of deferred items is determined by a user-defined parameter called coverage $c$, i.e., the percentage of cases for which the ML model provides the prediction without human assistance. E.g., if we aim for a target coverage of $90\%$, the total deferred cases should be approximately $10\%$ of all test instances.

\paragraph{Issues with LtD.}  It is well known that interaction with AI systems may trigger specific cognitive biases \citep{bertrand2022cognitive}, e.g., automation bias \citep{goddard2012automation, banerjee2024learning}, trust misplacement \citep{cabitza2025too}, and ordering effects \citep{nourani2021anchoring, pesenti2026human}, thus affecting -- and potentially impairing -- human decision making.

A handful of studies focusing specifically on the cognitive side of LtD exist.  \citet{hemmer2023human} showed that users can attain improved self-efficacy (and task satisfaction) when performing the target task on real deferred items, compared to items randomly extracted from the test set, suggesting that proper AI delegation \textit{can} improve human-AI team performance.
Other works portray a more nuanced picture.  \citet{bondi2022role} specifically studied how question framing impacts LtD, showing that presenting the model's prediction to the users induced an anchoring effect, which impacted their performance.  Building on this work, \citet{banerjee2024learning} argue that, besides potentially anchoring the human's judgment, the ML model provides no assistance on rejected cases, meaning the expert is left resolving the harder cases alone.
None of these works consider the consequences of algorithmic biases implicit in LtD strategies on users, nor describe the possible presence of the Test-taker's or Prevalence effects.

\paragraph{Test-taker's and Prevalence effects.}  The Test-taker’s effect is a psychological phenomenon, primarily studied in educational contexts \citep{lee2019test}, whereby test takers select answers that conform to their expectations about the overall answer distribution rather than relying on their own judgment, resulting in poorer performance. Such effect may stem from the students' previous test-taking experiences \citep{kiss2013gambler}, simple pattern recognition \citep{paul2014}, or reflect the representativeness heuristic \citep{tversky1974judgment}.
Normatively, students should consider the likelihood of each individual answer to be independent from the previous ones. However, as event streaks (e.g., the same answer repeated more than twice) are psychologically salient \citep{carlson2007rule}, they may break their
prior expectations that correct answers are positioned randomly, even when they arise from a genuinely random process. Such streaks may therefore create uncertainty or prompt test takers to revise their answers so that the resulting distribution more closely matches their expectations of a random sequence.

The Prevalence effect refers to the increased error rate observed in visual-search tasks when targets constitute no more than 10\% of the stimulus population \citep{wolfe2007low}.  It has been demonstrated in laboratory settings \citep{wolfe2010varying}, in simulations of real-world tasks involving trained personnel, such as airport-security screening \citep{wolfe2013prevalence}, and in real-world settings, such as mammographic cancer screening \citep{evans2013if}.
It results in an increase of omission errors (i.e., failures to detect the infrequent target) and shorter reaction times when the target is absent \citep{wolfe2005rare}. These outcomes have been attributed to a shift in the observer’s search criterion induced by the imbalanced distribution of stimuli.
These studies do carry over to LtD settings, due to the task nature, and the presence of an external AI agent.

While prior work has examined the algorithmic aspects of LtD and psychological phenomena arising both within and beyond human–AI interaction, to the best of our knowledge, we are the first to bring these perspectives together by investigating sampling biases in LtD strategies and assessing their direct impact on users interacting with these systems.

\begin{figure*}
    \centering
    \includegraphics[width=0.9\linewidth]{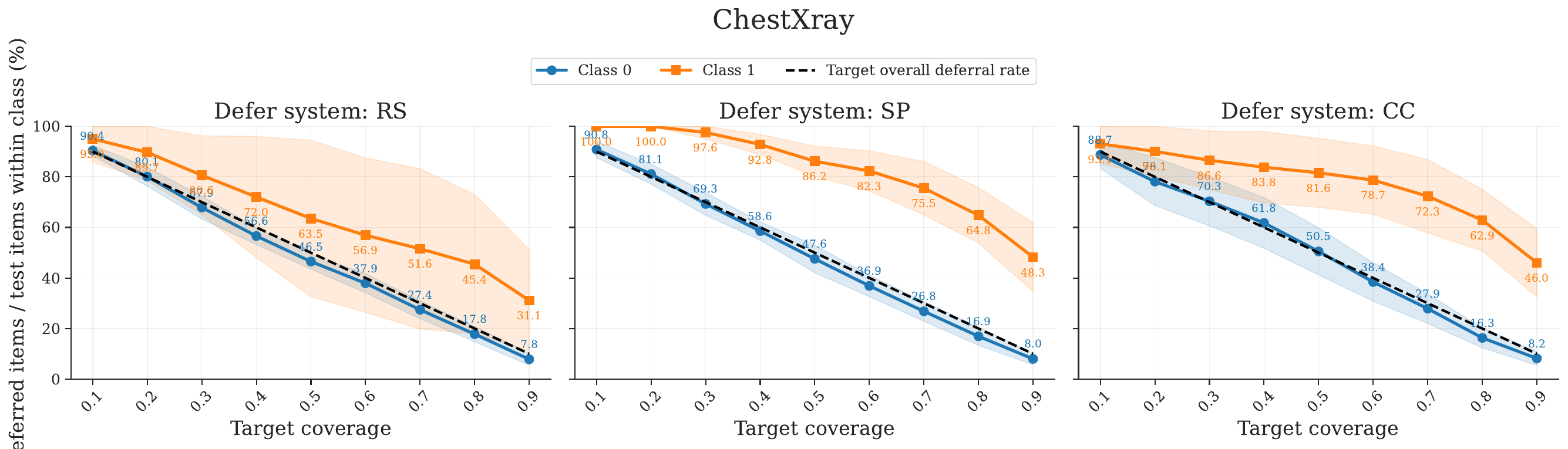}
    \caption{\textbf{LtD suffers from selective sampling in naturally imbalanced tasks}, here shown for \ChestXRay at different choices of target coverage ($x$-axis). The $y$-axis is the percentage of instances from each class that are deferred, with \textcolor{trainingcolor}{\textbf{blue}} indicating the majority class and \textcolor{testcolor}{\textbf{orange}} the minority class.  At all coverage levels, the minority class is disproportionately deferred.}
    \label{fig:natural-imbalance}
\end{figure*}

\section{Can LtD lead to Human Mistakes?}
\label{sec:issues}

LtD frameworks often assume the human expert to be a reliable and accurate agent for the delegated task \citep{DBLP:conf/aaai/RuggieriP25}.
Our work stems from the observation that \textit{human decision making can become inaccurate in settings that can trigger cognitive heuristics} \cite{tversky1974judgment, pohl2004cognitive}.
For instance, the expert might recognize patterns in the deferred cases (such as visual cues, common model's mistakes), which are present specifically in the most deferred class, allowing them to produce heuristics (i.e., inaccurate cognitive approximations) to solve the deferred cases with a lessened cognitive load, but at the expense of overall performance in cases in which the deferred items slightly differs from the usual, i.e., they belong to another class.
Intuitively, consider an airport checkpoint screener that receives deferred luggage scans. Depending on the specific rejection metric, the predictor tends to defer only false-positive cases, based on suspicious but ultimately innocuous pointy objects. The screener inadvertently produces a heuristic, whereby they quickly discard pointy objects as non-dangerous. But then a scan gets deferred with a similar, but this time dangerous, object; in this case, the screener quickly, but wrongly, rejects the screening, causing a misprediction.
Conversely, users might build wrongful expectations (such as a balanced number of deferred items for each class) about the distribution of deferred items by the model, and become wary once this expectation fails (e.g., when the model chooses to defer only one specific class), resulting in the adoption of inaccurate heuristics in favour of their own judgment.

We posit that existing LtD strategies may induce exactly this kind of issue, which can impair downstream decision-making, especially when the original training set presented few instances of a specific class.
Our first hypothesis is that:
\begin{quote}
    \textbf{Hypothesis 1} ($H_1$): \textit{LtD solutions can produce imbalanced rejection sets when the original training set is itself imbalanced with respect to a certain class.}
\end{quote}
We argue this can happen when the predictor is more confident about one class compared to the others. This makes the former more likely to be deferred.
Our second hypothesis is:
\begin{quote}
    \textbf{Hypothesis 2} ($H_2$): \textit{Imbalanced rejection sets can systematically affect human decision making.}
\end{quote}
We believe imbalanced rejection sets can influence human performance through two distinct cognitive paths: reduced attention and a mismatch between expected and actual class distributions.
The former is related to results from the visual search literature \citep{wolfe2005rare, wolfe2007low}, which show that infrequent search targets are less likely to be detected than more frequent ones.
The latter arises from the tendency of individuals to consider not only which answer they believe is correct, but also which answer best matches their expectations about the overall distribution of responses within the task (e.g., selecting option ``A'' rather than ``B'' because all previous answers were ``B''). This has been primarily studied in the context of multiple-choice test design \citep{lee2019test, paul2014}.

In the following, we produce evidence in support of both hypotheses.
In \cref{sec:synthetic-experiments} we focus on $H_1$ and study empirically the relationship between training class imbalance and rejection set bias induced by representative LtD strategies.
In \cref{sec:user-study}, we introduce a larger-scale user study linking selection bias to cognitive biases, allowing us to ground $H_2$.

\section{Learning to Defer Induces Sampling Bias}
\label{sec:synthetic-experiments}

In this section, we evaluate whether representative LtD algorithms induce sampling bias ($H_1$).
To this end, we rely on the LtD implementations by \citet{mozannar2020consistent} and \citet{palomba2024causal}, monitoring the composition of the rejection set when manipulating the percentage of majority-class items in the training set.

\begin{figure*}[!t]
    \centering
    \includegraphics[width=0.96\linewidth]{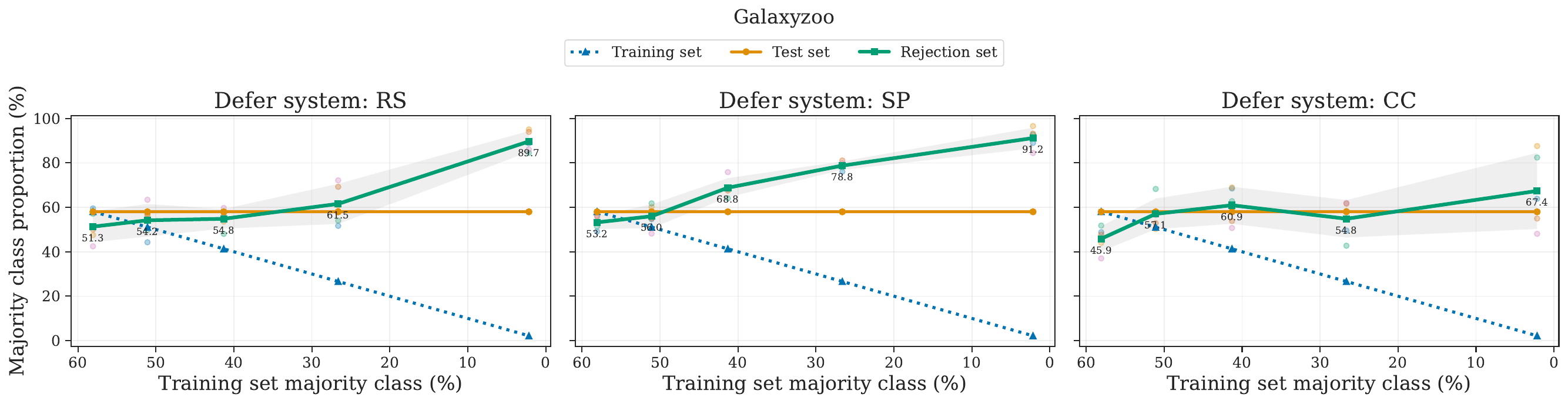}
    \includegraphics[width=0.96\linewidth]{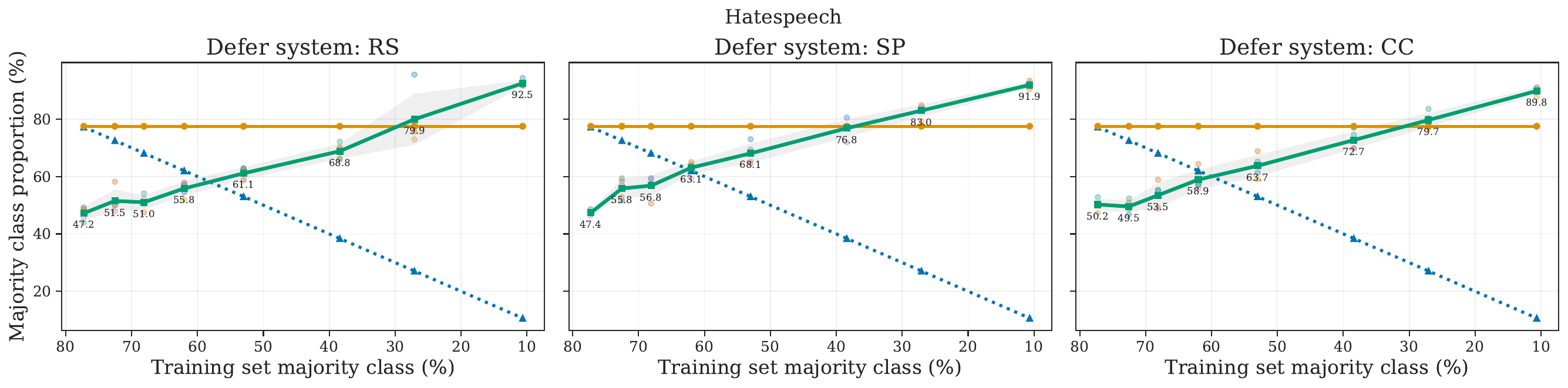}
    \includegraphics[width=0.96\linewidth]{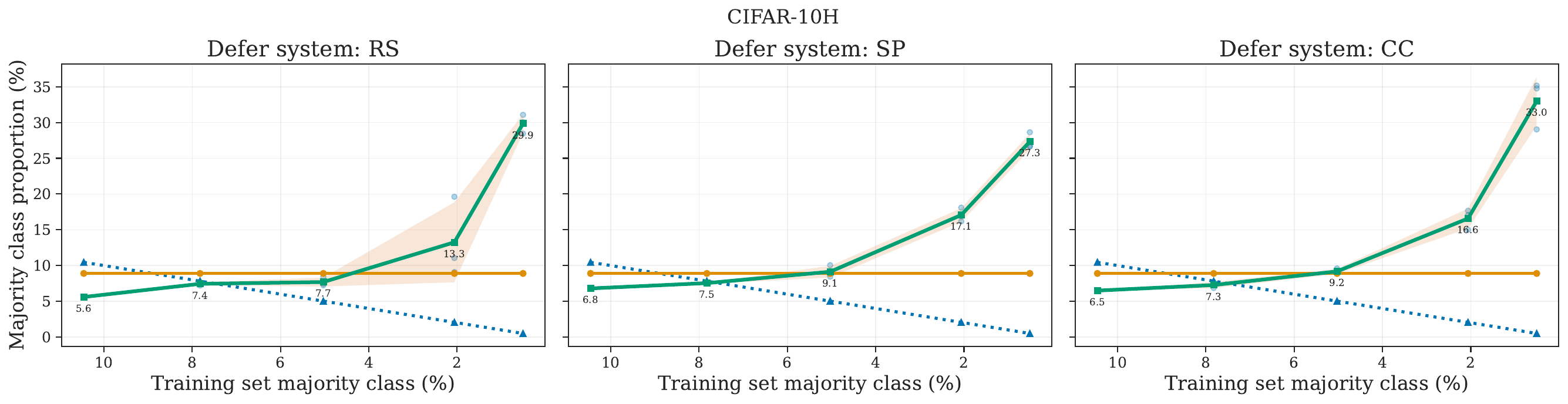}
    \caption{\textbf{Selective sampling can be induced via ablation}. Percentage of the (ablated) majority class in the training (\textbf{\textcolor{trainingcolor}{blue}}), test (\textbf{\textcolor{testcolor}{orange}}) and rejection (\textbf{\textcolor{deferredcolor}{green}}) set in \GalaxyZoo, \HateSpeech and \CifarH for all three LtD Strategies (SP, CC and RS).  The rarer the ablated class in the training set, the more it dominates the rejection set.}
    \label{fig:ablation}
\end{figure*}

\paragraph{Baselines.}  We evaluate three popular and state-of-the-art LtD strategies:
\textit{i}) \textit{\underline{Selective Prediction}} (SP) \citep{geifman2017selective} defers predictions based on the classifier's uncertainty. More precisely, the method uses the maximum of the predicted probabilities as a confidence score.
\textit{ii}) \textit{\underline{Com}p\underline{are Con}f\underline{idence}} (CC) \citep{raghu2019algorithmic} jointly trains a classifier and a human-error model. The method then provides a confidence score that is the difference between the confidence of the classifier and the human-error model.
\textit{iii}) \textit{\underline{Realizable Surro}g\underline{ate}} (RS) \citep{mozannar2023should}, a state-of-the-art LtD model that trains both the ML model and the score used for deferral using an ad-hoc loss.

Following standard practice in LtD~\citep{mozannar2023should}, each method defers items by comparing the respective confidence score with a threshold $\tau$. This threshold is estimated on a held-out set to achieve the desired target coverage.

\paragraph{Data sets and classifiers.}  We consider four real-world human-annotated datasets, following the same processing protocol and training setup as \citet{palomba2024causal}.
{\tt \underline{ChestXRa}y} \citep{majkowska2020chest} consists of chest X-ray scans with expert humans annotations over four pathological conditions (fracture, pneumothorax, airspace opacity, and nodule or mass) and a pre-defined ground truth label.  The task was binarized, taking ``pneumothorax'' to be the positive class.
{\tt \underline{Galax}y\underline{Zoo}} \citep{fortson2012galaxy} is a binary task about classifying images of galaxies as ``smooth'' or ``non-smooth'', each associated with $30$ human predictions for the task and a majority-vote ground-truth label.
{\tt \underline{HateS}p\underline{eech}} \citep{davidson2017automated} contains $25$K human annotated tweets, and the classification task consists in categorizing a tweet into one of three classes: ``hatespeech'', ``offensive language but not hatespeech'', or ``neither''. Each tweet comes with $3$ human annotations and a majority-vote ground-truth label.
\underline{\CifarH} \citep{peterson2019humanuncertaintymakesclassification, krizhevsky2009learning} contains $60$K images belonging to $10$ categories (airplane, automobile, bird, cat, deer, dog, frog, horse, ship, truck). The $10$K test set images also contain human predictions and the original label as the ground truth. 
Following \citet{mozannar2023should} and \citet{palomba2024causal}, for \ChestXRay and \GalaxyZoo we fine-tuned a pre-trained DenseNet121 \citep{huang2017densely}, restricting fine-tuning to the non-human-annotated images for \ChestXRay, while for \HateSpeech we fine-tuned a logistic regressor on top of {\tt SBERT} embeddings and for \CifarH we fine-tuned a WideResNet \citep{zagoruyko2016wide}. For all datasets, we employed the {\tt Adam} or {\tt AdamW} optimizers (\ChestXRay: $10$ epochs / learning rate $0.0001$; \GalaxyZoo: $50$/$0.001$; \HateSpeech: $100$/$0.01$; \CifarH: $200$/$0.001$) with a uniform batch size of $128$ and a 70-20-10 train-test-validation split.

\paragraph{Results.}  We examined \ChestXRay without artificially ablating its training set. The dataset is naturally imbalanced, with the minority class constituting only $3.21\%$ of the total training instances. We report in \cref{fig:natural-imbalance}, for each target coverage, on the X axis, the percentage deferred instances of each class on the total number of instances in the test set, on the Y axis. As shown in the plots, for all deferral strategies, the minority class (\textcolor{testcolor}{\textbf{orange}} line) gets disporportionately more deferred than the majority class, with up to $100\%$ of deferred instances at low coverage, and, for RS, SP, and CC, respectively, $38.1\%$, $48.3\%$, and $46\%$ at coverage 0.9.
\cref{fig:natural-imbalance} shows that the minority class is overrepresented across all target coverage levels, with the degree of overrepresentation increasing at higher coverage levels.

Then, for all other datasets, we manipulated the percentage of majority class items in the training set, each time retraining the classifier as described above and the rejector following \citep{palomba2024causal} verbatim, at a target coverage at $0.9$ -- i.e., $10\%$ of the cases are deferred, chosen to simulate situations in which the cost of deferral is high -- and then measured the proportion of the ablated class in the rejection set.  We report classification performance, for transparency, in Table 1 in the Supplementary Materials.

Our results are reported in \cref{fig:ablation}.
Ablation was effective across all methods and defer strategies.
For \GalaxyZoo, ablation was effective on RS and SP, with majority class proportion going from, respectively, $51,3\%$ and $53.2\%$, to $89.7\%$ and $91,2$; while CC was the most insensitive, with the majority class going from $45,9\%$ to $67,4\%$
For \CifarH, all of the defer systems relatively suffered the dataset ablation, with class proportions going from $5.6\%$ to $29.9\%$ for RS, from $6.8\%$ to $27.3\%$ for SP, and from $6.5\%$ to $33.0\%$ for CC. 
Similarly, for \HateSpeech, all of the defer system suffered the ablation, with class proportions going from $47.2\%$ to $92.5\%$ for RS, from $47.4\%$ to $91.9\%$ for SP, and from $50.2\%$ to $89.8\%$ for CC.

Overall, these results support ($H_1$), showing that, when trained on either naturally or artificially imbalanced datasets, LtD strategies preferentially defer instances of the class underrepresented in the training set.

\section{Sampling Bias Leads to Human Mistakes}
\label{sec:user-study}

We test $H_2$ by evaluating the possible detrimental impact of LtD
on users' performance using the \GalaxyZoo dataset \citep{fortson2012galaxy}.  We specifically chose this dataset since its binary classification task does not require domain expertise (given that the dataset is annotated through crowdsourcing), and reduces the need for subjective interpretation of the stimuli, eliminating one potential source of noise.

\begin{figure}[!t]
    \centering
    \includegraphics[width=0.75\linewidth]{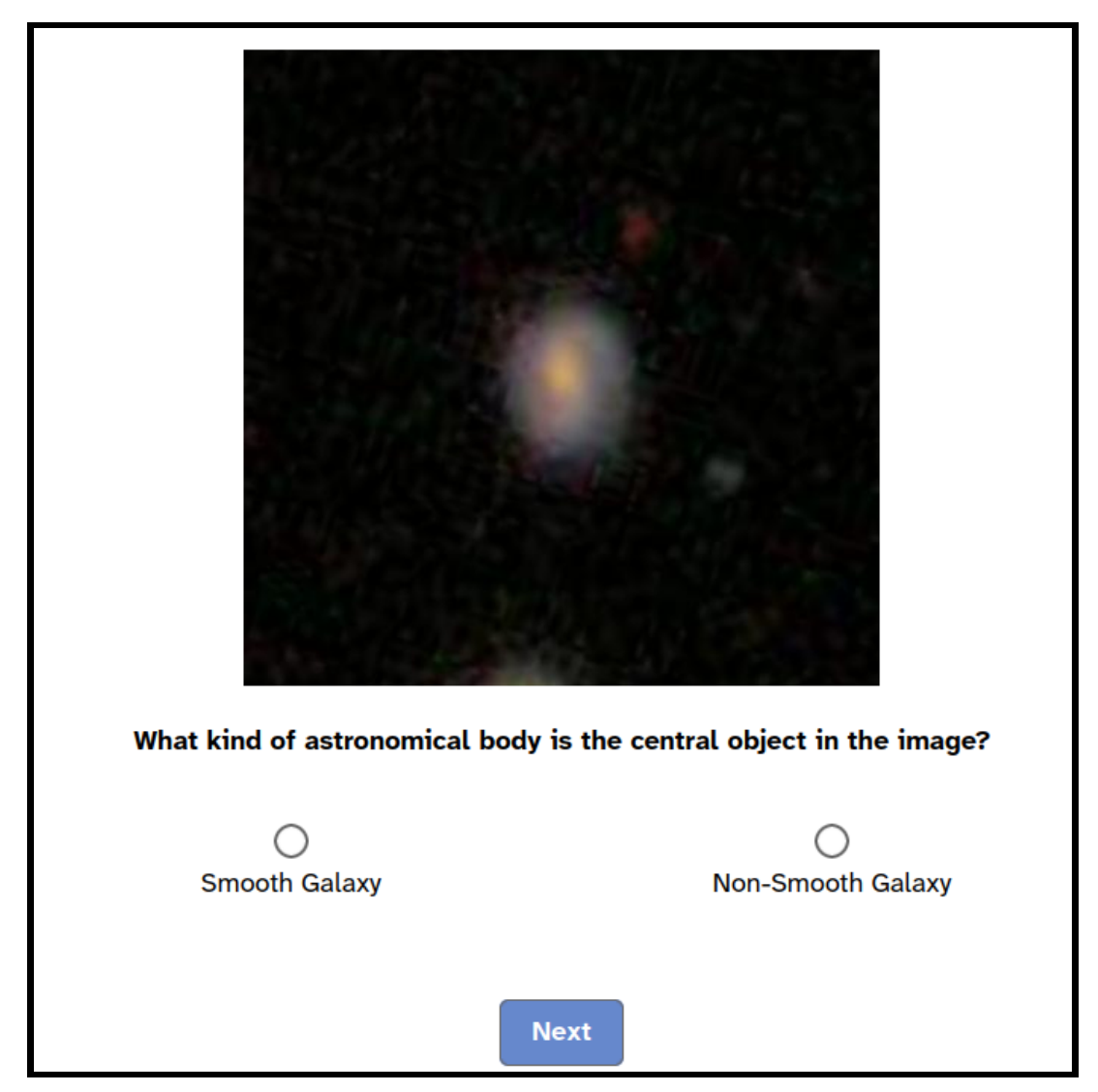} 
    \caption{\textbf{Experimental interface} showing an example smooth galaxy from \GalaxyZoo and the two alternative classes that participants can choose from.}
    \label{fig:interface}
\end{figure}

\paragraph{Task and data.}  We implemented a web interface for \GalaxyZoo (see \cref{fig:interface}).
Participants were exposed to 150 images of galaxies that the RS strategy deferred, and they had to classify each of them as belonging to one of two classes, ``smooth'' and ``non-smooth''. We chose this number not to cause fatigue effects, but still grant statistical power.
We reduced variability attributable to the stimuli themselves, by manually vetting low-quality images, images presented from unusual viewing angles, or exhibiting atypical characteristics.
Additionally, we excluded galaxies viewed edge-on with a galaxy-morphology classifier \citep{semenov2025galaxy}; then we removed the images with an annotator agreement lower than $60\%$ in the original dataset, finally, we manually selected a sample of $135$ images for each class.

\paragraph{Variables and analyses.}  We manipulated the following two independent variables:
\begin{itemize}
    \item \textit{Deferred condition}: it refers to the proportion of smooth/non-smooth galaxies presented to the participant and comprised three conditions, namely \textit{balanced}, with a $50\%$-$50\%$ split of smooth and non-smooth images, which represented the control condition; \textit{smooth majority}, with a $90\%$-$10\%$ split, and \textit{non-smooth majority} with a $10\%$-$90\%$ split. We chose these proportions to reflect previous literature in visual search tasks \citep{wolfe2013prevalence, wolfe2010varying}. This variable was manipulated between-subjects, meaning that each participants was assigned to only one of the three conditions.
    \item \textit{Stimulus ground-truth}: it refers to the class of stimuli, determined based on the class annotated by the majority of the original dataset annotators. It is comprised of two levels, \textit{smooth} and \textit{non-smooth}, and was treated as a within-subjects variable, meaning that participants were exposed to both of its levels.
\end{itemize}
To minimize stimulus-related confounds, we used the same images across all experimental conditions. Specifically, all smooth and non-smooth images presented in each condition were sampled from the same two pools of $135$ images described above, with only the sampling proportions varying across conditions.
Specifically, in the smooth-majority condition, all $135$ smooth images were presented (i.e.,, $90\%$ of the $150$ deferred images), whereas the remaining $15$ non-smooth images ($10\%$) were randomly sampled from the pool of $135$ non-smooth images. Conversely, in the non-smooth-majority condition, all $135$ non-smooth images were presented and 15 smooth images were randomly sampled. Finally, in the balanced condition, $75$ images ($50\%$) from each class were randomly sampled from their respective pools.

\paragraph{Participants and data exclusions.} An a priori power analysis was performed using {\tt SimR} \citep{green2016simr, kumle2021estimating}, which estimated a sample size of $210$ to achieve a power of 84\% to detect a small effect size of the interaction between deferred condition and stimulus ground-truth on participants' classification accuracy.
A total of 300 participants were recruited on Prolific (www.prolific.com), among native English spearkers with a Prolific approval rate of at least 95\%, and compensated in accordance with the hourly rate suggested by Prolific: 1.2£ plus a possible bonus of 5£ for the 15 participants that scored the highest performance.
We excluded from the analyses participants who left the experiment browser tab more than once during the experiment, resulting in 226 participants retained\footnote{Including these participants produced only minimal differences in the results (reported in the Results section), none of which altered their overall interpretation. Applying a more stringent exclusion criterion (excluding participants who left the browser tab even once) reduced the final sample below the size indicated by the power analysis; however, the pattern of results remained unchanged.}, $M_{Age} =  39.26 \pm 12.34$\footnote{When reporting means, we refer to the raw averages, and values after $\pm$ denote the Standard Deviation, while when reporting model's results and comparisons, it denotes Standard Error}, 53$\%$ males, with a balanced distribution across conditions: $N_{\textit{Balanced}} = 73$; $N_{\textit{Smooth}} = 72$; $N_{\textit{Non-smooth}} = 81$.

\paragraph{Procedure.} After providing informed consent, participants were instructed on the task (including being informed on the performance-based bonus to motivate attentive responding) and the definition of smooth and non-smooth galaxies (adapted from the \GalaxyZoo online project; \citealt{fortson2012galaxy}).\footnote{The instructions did not differ across conditions.  The exact wording of the task instructions and the questionnaire items are reported in the Supplementary Material.} They were then shown $6$ example images for each class, then completed $4$ warm up trials, without providing feedback to participants.
Participants were then randomly assigned to one of the three deferred-set conditions and completed the $150$ classification trials, with the proportion of smooth and non-smooth galaxies determined by their assigned condition. The order of image presentation was randomized independently for each participant. Importantly, no feedback on classification accuracy was provided during the task. This choice aimed to simulate a realistic annotation setting, where ground-truth labels may not be immediately available, and to limit learning effects across trials.

\begin{figure*}[!t]
    \centering
    \includegraphics[width=0.88\linewidth]{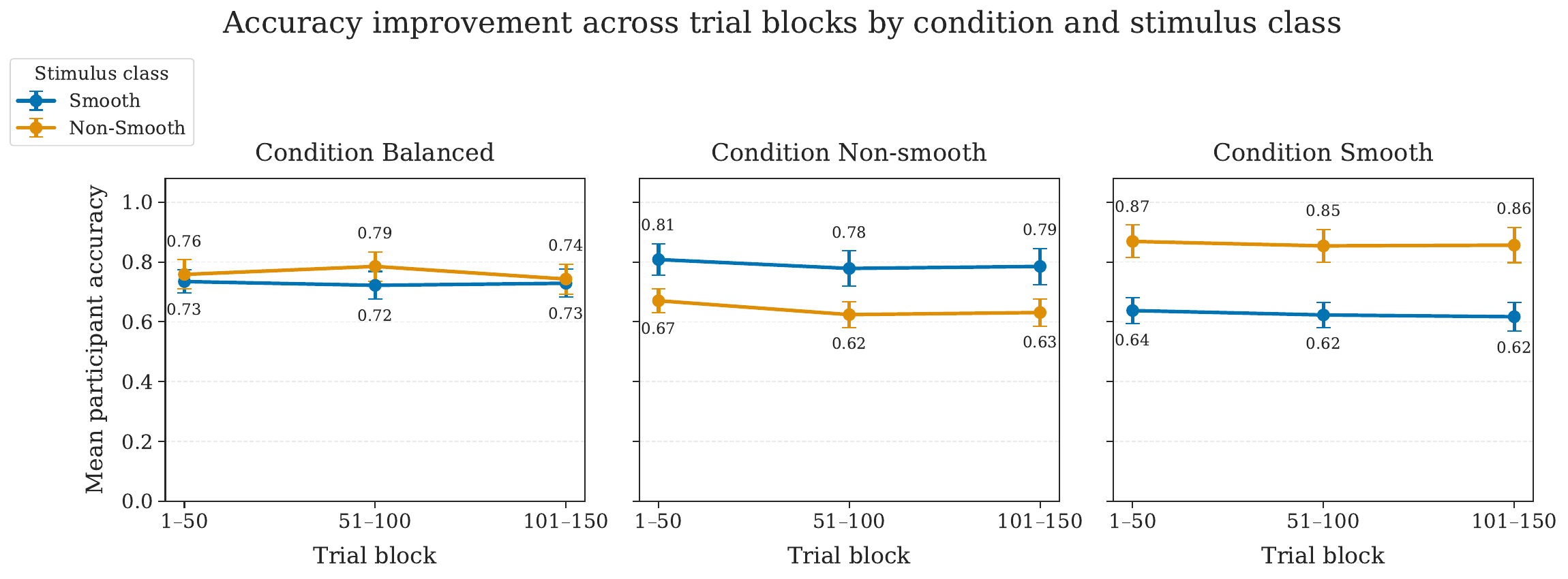}
    \caption{\textbf{Participants performed worse on whichever class
    is the majority in their respective condition}, as shown by the average accuracy across time per class plotted above.
    No training effect was found in any condition.}
    \label{fig:main-results}
\end{figure*}

\paragraph{Results.} Participants' classification accuracy was analyzed using a logistic mixed-effect regression model. The model included experimental condition and ground-truth class, along with their interaction, as fixed effects, as well as random intercepts for participant and Image ID.
The effect of condition was not significant ($\chi^2(2) = 0.62$, $p=.735$ $M_{Balanced}=0.75 \pm 0.44$, $M_{Smooth}=0.65\pm0.48$, $M_{Nonsmooth}=0.66\pm0.49$), and neither was the effect of stimulus class ($\chi^2(1)=0.29$, $p=.590$, \textit{Smooth}: $0.67 \pm .47$, \textit{Non-Smooth}: $0.69 \pm 0.46$). 
However, the interaction between condition and stimulus ground-truth was significant ($\chi^2(2)=278.95$, $p<.001$).
Post-hoc pairwise comparisons (\textit{p} values are Bonferroni-corrected for a total of 6 contrasts to maintain the family-wise error rate at $\alpha = .05$) indicated that, for smooth images, participants in the \textit{Smooth majority} condition performed worse ($M=0.62\pm0.48$)  than participants in the \textit{Balanced} ($M=0.72\pm0.44$, Odds Ratio: $2.19 \pm 0.33, z = 5.07, p < 0.001$) and \textit{Non-smooth majority} condition ($M=0.79 \pm 0.40$, OR: $3.28 \pm 0.57, z = 6.86, p < .001$), while there were no significant differences between \textit{Balanced} and \textit{Non-smooth majority} (OR: $0.66 \pm 0.11, z = -2.31, p = .123$).\footnote{the latter comparison becomes significant when not excluding any participant (OR: $0.62 \pm 0.10, z=-3.11, p=.011$).}
Conversely, when classifying non-smooth images, participants in the \textit{Smooth majority} condition were significantly more accurate ($M=0.86\pm0.35$) than participants in the \textit{Balanced} ($M=0.76 \pm 0.43$, OR: $1,85, z=3.29, p=.006$) and \textit{Non-Smooth majority} condition ($M= 0.64\pm0.48$, OR:$3.47\pm0.63, z=6.89, p<.001$), while there was also a significant difference in accuracy between the \textit{Balanced} and \textit{Non-smooth majority} condition ($1.89\pm0.28, z=4.23, p<.001$).

Additionally, as shown in \cref{fig:main-results}, the average accuracy did not change over time in any condition or class of stimulus, meaning that the effect we found is unlikely to be explained by lower familiarization at the beginning of the task or fatigue towards its end. We ran an additional logistic
mixed-effect regression model keeping accuracy as dependent variable, using experimental condition and trial block (defined as consecutive number of $50$ items block) as fixed effects. Results were not significant for trial block ($\chi^{2}=0.80, p=.671$) nor were significant for the interaction ($\chi^{2}=3.98, p=.408$).

\begin{figure}[!t]
    \centering
    \includegraphics[width=0.9\linewidth]{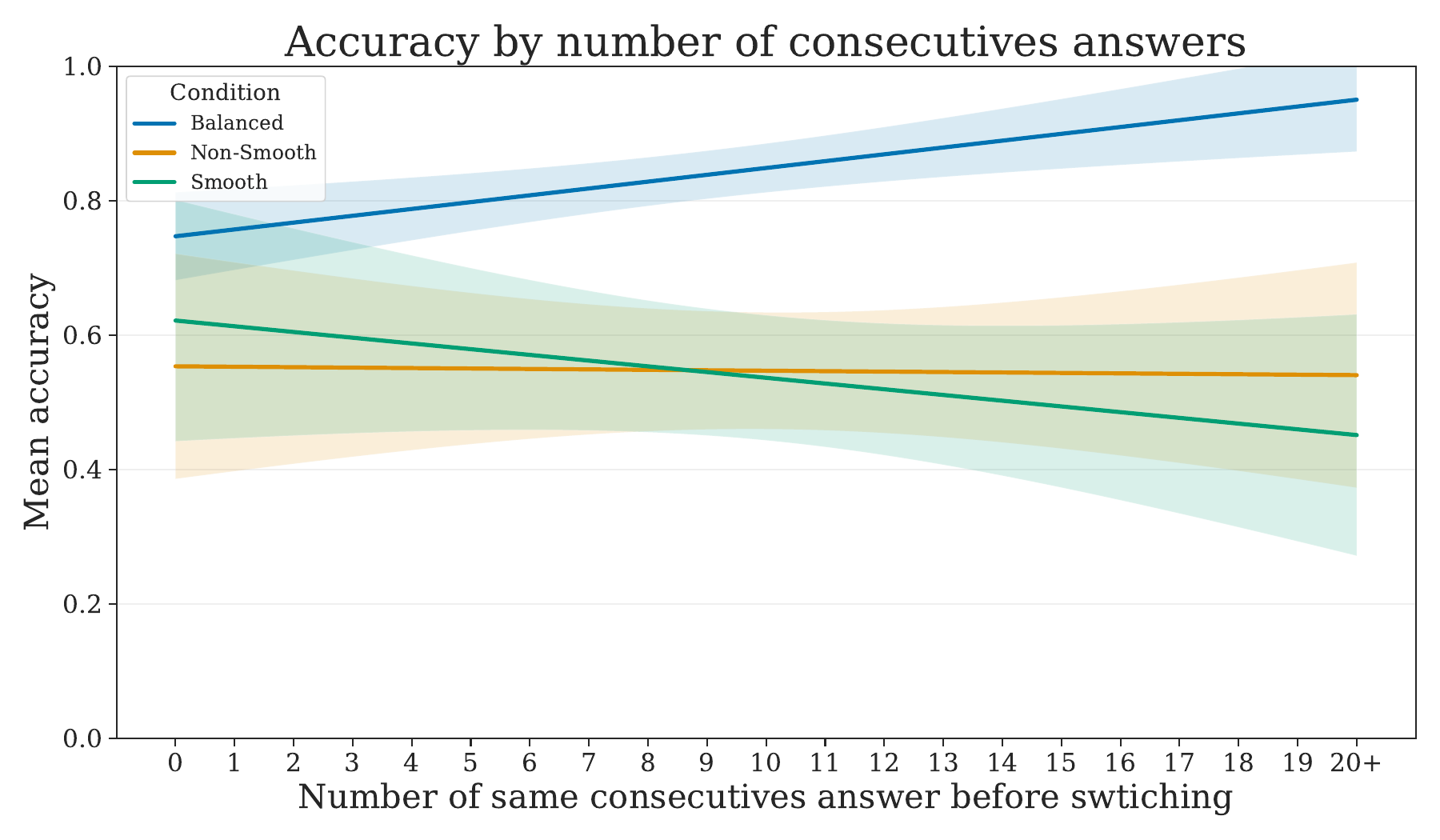}
    \caption{\textbf{In the imbalanced conditions, accuracy when switching declines after longer streaks of identical responses}. 
    On the $x$-axis: number of consecutive times participants gave the same answer before switching. On the $y$-axis: average accuracy when switching. }
    \label{fig:consecutives}
\end{figure}

\paragraph{Evidence for Test-taker's effect.}  Our results do not appear to support an attentional-bias account, as this would predict the largest decrease in accuracy for the minority class. Instead, they appear to be more consistent with a Test-taker's effect.  These analyses aimed at assessing whether, when participants switched their classification compared to the one selected in the preceding trial, they did so correctly or not, and whether higher numbers of consecutive identical answers increased the probability of performing incorrect switches (i.e., switches leading to an incorrect classification).
To explore this possibility, we conducted a generalized linear mixed-effects model only on the trials in which participants changed answer from their previous response, with accuracy as dependent variable, condition and the number of consecutive identical responses a participant gave before changing their answer as fixed effects, and keeping Participant and Image ID as random effects.

The main effect of consecutive answers was not significant ($\chi^2(1)=1.95, p=.163$), while effect of condition was ($\chi^2(2)=87.02, p<.001$), as well as their interaction ($\chi^2(2)=15.62, p<.001$).
Post hoc comparisons of the linear trends across the three deferral conditions revealed that the regression coefficient for the number of consecutive identical responses was significantly different in the \textit{Balanced} condition than in both the \textit{Smooth-majority} condition (difference: $0.11 \pm 0.03$, $z = 3.89$, $p < .001$) and the \textit{Non-smooth-majority} condition (difference: $0.08 \pm 0.03$, $z = 3.31$, $p = .003$). In contrast, the regression coefficients did not differ significantly between the \textit{Smooth-majority} and \textit{Non-smooth-majority} conditions (difference: $0.02 \pm 0.02$, $z = 1.10$, $p = .818$). Specifically, inspecting the regression coefficients in the three conditions reveals the number of consecutive identical answers had a negative impact on participants' accuracy in the \textit{Smooth majority} ($-0.08\pm0.02$) and \textit{Non-smooth majority} ($-0.05\pm0.01$) conditions, but not in the \textit{Balanced} ($0.03\pm0.02$) one (see \cref{fig:consecutives}).

\section{Discussion and Conclusion}
\label{sec:discussion}

Overall, our findings suggest that LtD can affect human accuracy ($H_2$) by shaping the distributions of deferred cases ($H_1$).
Our ablation experiments (\cref{sec:synthetic-experiments}) show that training-time class imbalance influences the composition of the rejection set for all LtD strategies we tested.

At the same time, our user study (\cref{sec:user-study}) suggests that, in such cases, participants are less accurate on whichever class constitutes the majority of their assigned cases.
Exploratory analyses further suggest that this pattern may reflect a Test-taker’s effect: after longer streaks of identical responses, participants in the imbalanced conditions were more likely to switch incorrectly than those in the balanced condition.
While these results cannot be considered conclusive to the assessment of this cognitive bias, they demonstrate a clear response pattern distinct from Prevalence effects, which would instead be proven by the opposite pattern (i.e., users being less accurate on the \textit{minority} class).

More broadly, these results challenge the LtD assumption that human performance is invariant to the deferral policy. Reliable LtD systems should therefore account not only for which cases are deferred, but also for how the resulting rejection-set distribution affects human decision making. Potential mitigations include informing users that deferred cases may be non-uniformly distributed and developing deferral frameworks that explicitly model such distribution-dependent changes in human performance.

\section*{Ethical Statement}

Our study has received approval from the Ethics board of our university, document identifier code 2026-028ESA.

\section*{Acknowledgments}
Funded by the European Union. Views and opinions expressed are however those of the author(s) only and do not necessarily reflect those of the European Union or the European Health and Digital Executive Agency (HaDEA). Neither the European Union nor the granting authority can be held responsible for them. Grant Agreement no. 101120763 - TANGO.

\newpage

\bibliographystyle{plainnat}
\bibliography{references}
\newpage
\appendix

\section{LtD Experimental Details}

\begin{table}[!h]
    \caption{Machine-learning accuracy at target coverage $1$.}
    \label{tab:accuracy_coverage_04}
    \centering
    \scalebox{0.85}{
    \begin{tabular}{lcc}
        
        {\sc Dataset} & {\sc Method} & {\sc Accuracy} \\
        \midrule
        \GalaxyZoo 
        & RS & 0.82 $\pm$ 0.011 \\
        & SP & 0.82 $\pm$ 0.009 \\
        & CC & 0.82 $\pm$ 0.009 \\
        \midrule
        \HateSpeech 
        & RS & 0.87 $\pm$ 0.002 \\
        & SP & 0.88 $\pm$ 0.005 \\
        & CC & 0.88 $\pm$ 0.004 \\
        \midrule
        \CifarH 
        & RS & 0.92 $\pm$ 0.007 \\
        & SP & 0.92 $\pm$ 0.007 \\
        & CC & 0.92 $\pm$ 0.008 \\
        \bottomrule 
    \end{tabular}
    }
\end{table}

\section{User Study Instructions}
\subsection{Page 1}
Welcome to the study

In this study you will be presented with images of galaxies and asked to classify them as either smooth or non-smooth.

The study will take approximately 10 minutes to complete.

Your participation is voluntary, and you are free to withdraw from the study at any time without providing a reason. Please note that if you choose to withdraw, you will not receive any compensation.

If you have any questions regarding the study or the handling of your data, please contact us at ANONYMIZED.

To proceed, we require your informed consent.

\subsection{Page 2}

An \textbf{AI model} was trained to classify telescope images of
\textbf{galaxies} as either \textbf{smooth} or \textbf{non-smooth}, and was
subsequently deployed on a set of galaxy images.

For 150 images, the model decided it was safer/better to involve a human
decision maker rather than directly predicting. You will therefore be
presented with these 150 images, one at a time. Your task is to classify each
galaxy as either smooth or non-smooth based on the image shown.
\textbf{How to classify galaxies}
\newpage
\noindent\textbf{Smooth} galaxies gradually \textbf{fade in all directions} from the
center. It is often hard to pinpoint an edge of a smooth galaxy. There may be
a \textbf{small, bright, symmetric core}.
\begin{figure}[H]
    \centering
    \begin{tabular}{@{}ccc@{}}
        \includegraphics[width=0.28\columnwidth]
        {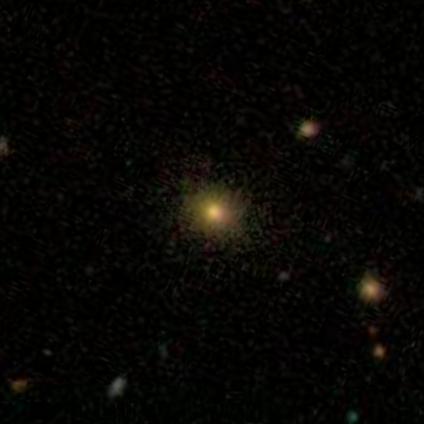}
        &
        \includegraphics[width=0.28\columnwidth]
        {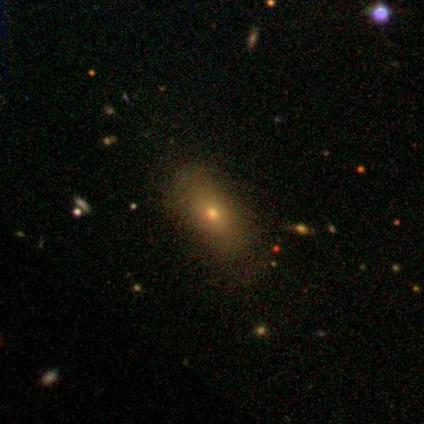}
        &
        \includegraphics[width=0.28\columnwidth]
        {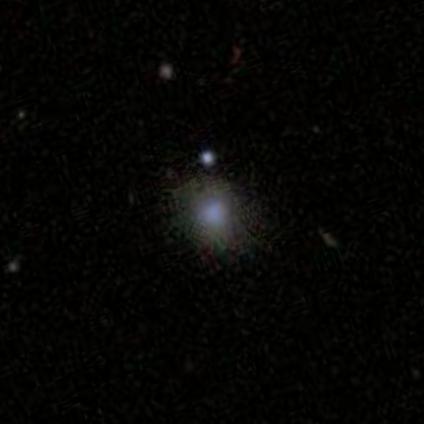}
        \\[4pt]
        \includegraphics[width=0.28\columnwidth]
        {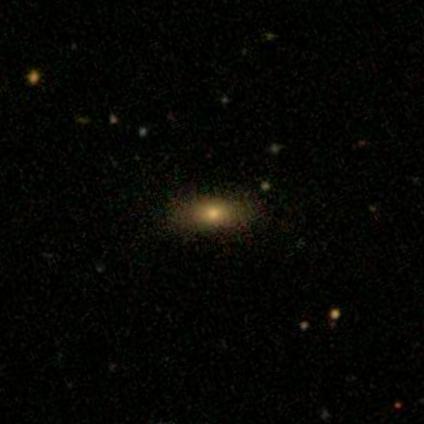}
        &
        \includegraphics[width=0.28\columnwidth]
        {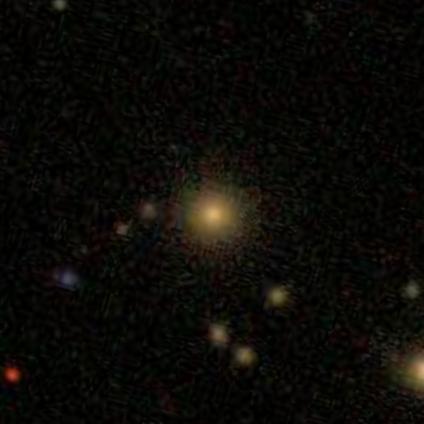}
        &
        \includegraphics[width=0.28\columnwidth]
        {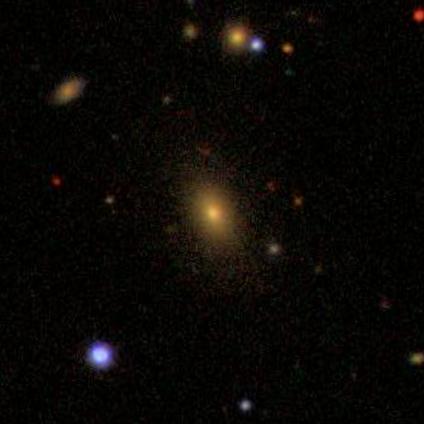}
    \end{tabular}
    \label{fig:smooth-examples}
\end{figure}

\noindent\textbf{Non-smooth galaxies} show visible structure or features. These may
include \textbf{spiral arms}, a \textbf{bulge}, a \textbf{bar}, or any other
\textbf{irregular or distinct features}.

\begin{figure}[H]
    \centering
    \begin{tabular}{@{}ccc@{}}
        \includegraphics[width=0.28\columnwidth]
        {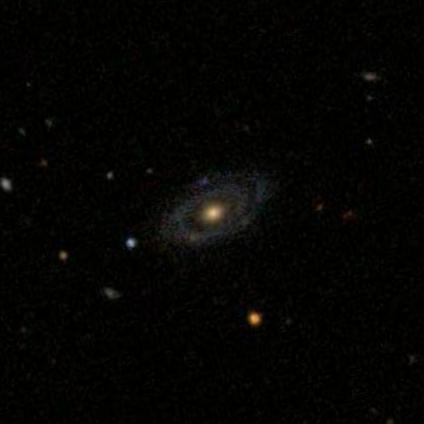}
        &
        \includegraphics[width=0.28\columnwidth]
        {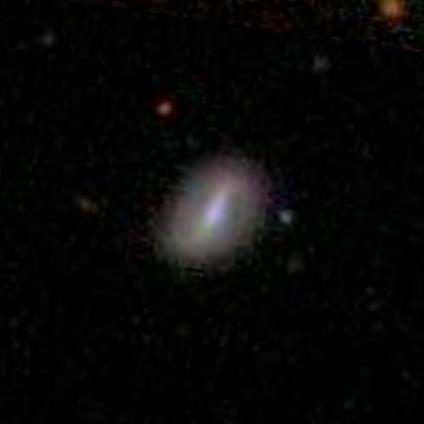}
        &
        \includegraphics[width=0.28\columnwidth]
        {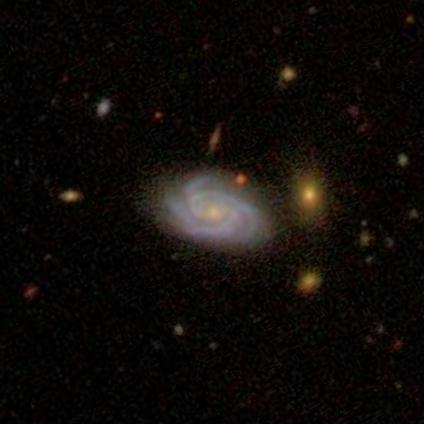}
        \\[4pt]
        \includegraphics[width=0.28\columnwidth]
        {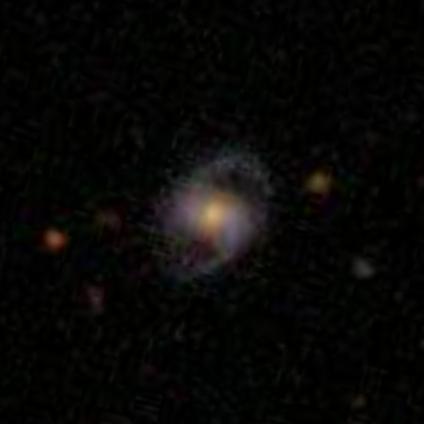}
        &
        \includegraphics[width=0.28\columnwidth]
        {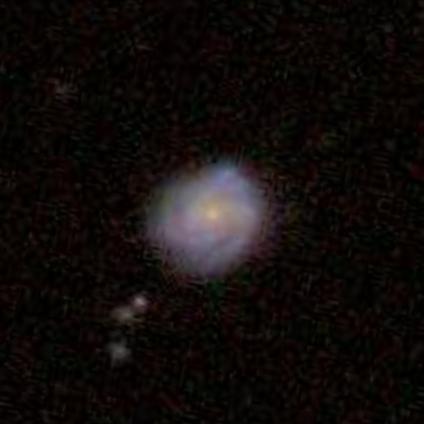}
        &
        \includegraphics[width=0.28\columnwidth]
        {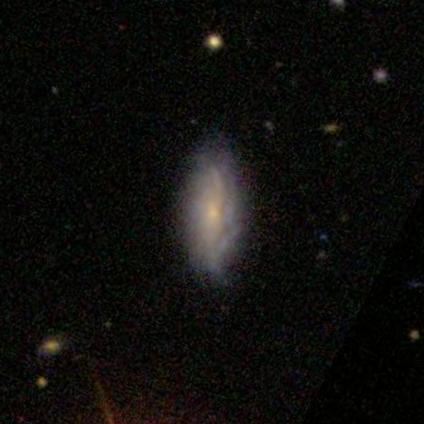}
    \end{tabular}
    \label{fig:non-smooth-examples}
\end{figure}
\textbf{Important:} For each image shown, \textbf{classify only the
galaxy located at the centre of the image}. Ignore any other objects, galaxies,
stars, or artifacts that may appear elsewhere in the image.

\textbf{Before the main task}

\noindent Before you begin classifying the 150 images, you will be shown example images
of smooth and non-smooth galaxies. You will then complete four practice trials.

\textbf{Important}

\noindent Try to be as accurate as possible. The five participants with the highest
accuracy will receive a bonus of \pounds 2.

\noindent To ensure data quality, it is important that you \textbf{remain on the study
page for the entire duration of the task}. Participants who do not follow the
instructions may be \textbf{excluded from future studies conducted by our
research group}.

\newpage
\section{User study results}
In the following tables, all predictors were deviation coded, reference levels are reported between brackets. For post-hoc comparison, Bonferroni-corrected p-values are reported. SE denotes Standard Error, OR denotes Odds Ratio.
\begin{table*}[htbp]
    \centering
    \caption{Results of the main logistic linear mixed-effects model predicting trial-level accuracy, and post-hoc contrasts.}
    \label{tab:main-glmm-results}
    \small
    \begin{tabular}{llrrrr}
        \toprule

        \multicolumn{6}{l}{\textit{Regression coefficients}} \\
        \addlinespace[2pt]
        \multicolumn{2}{l}{Fixed effect}
            & Estimate & SE & $z$ & $p$ \\
        \midrule
        \multicolumn{2}{l}{Intercept}
            & 1.36 & 0.08 & 16.24 & $<.001$ \\
        \multicolumn{2}{l}{Condition 1 (Smooth majority)}
            & 0.07 & 0.09 & 0.76 & .448 \\
        \multicolumn{2}{l}{Condition 2 (Smooth majority)}
            & $-0.05$ & 0.09 & $-0.53$ & .599 \\
        \multicolumn{2}{l}{Ground truth 1 (Non-smooth galaxy)}
            & $-0.03$ & 0.06 & $-0.54$ & .590 \\
        \multicolumn{2}{l}{Condition 1 $\times$ Ground truth 1}
            & 0.06 & 0.03 & 1.97 & .049 \\
        \multicolumn{2}{l}{Condition 2 $\times$ Ground truth 1}
            & 0.58 & 0.04 & 15.25 & $<.001$ \\

        \addlinespace[6pt]
        \multicolumn{6}{l}{\textit{ANOVA omnibus tests}} \\
        \addlinespace[2pt]
        \multicolumn{2}{l}{Fixed effect}
            & $\chi^2$ & $df$ & & $p$ \\
        \midrule
        \multicolumn{2}{l}{Intercept}
            & 263.72 & 1 & & $<.001$ \\
        \multicolumn{2}{l}{Condition}
            & 0.62 & 2 & & .735 \\
        \multicolumn{2}{l}{Ground truth}
            & 0.29 & 1 & & .590 \\
        \multicolumn{2}{l}{Condition $\times$ Ground truth}
            & 278.95 & 2 & & $<.001$ \\

        \addlinespace[6pt]
        \multicolumn{6}{l}{\textit{Post-hoc contrasts}} \\
        \addlinespace[2pt]
        Ground truth & Condition contrast
            & OR & SE & $z$ & $p$ \\
        \midrule
        Smooth
            & Balanced / Non-smooth
            & 0.67 & 0.12 & $-2.32$ & .124 \\
        Smooth
            & Balanced / Smooth
            & 2.19 & 0.34 & 5.07 & $<.001$ \\
        Smooth
            & Non-smooth / Smooth
            & 3.28 & 0.57 & 6.86 & $<.001$ \\
        \addlinespace
        Non-smooth
            & Balanced / Non-smooth
            & 1.89 & 0.28 & 4.23 & $<.001$ \\
        Non-smooth
            & Balanced / Smooth
            & 0.54 & 0.10 & $-3.29$ & .006 \\
        Non-smooth
            & Non-smooth / Smooth
            & 0.29 & 0.05 & $-6.89$ & $<.001$ \\
        \bottomrule
    \end{tabular}

    \vspace{0.5em}
    \begin{minipage}{0.95\linewidth}
        \footnotesize
        
    \end{minipage}
\end{table*}

\begin{table*}[htbp]
    \centering
    \caption{Results of the logistic linear mixed-effects model by retaining participants the never left the experiment tab, and post-hoc contrasts.}
    \label{tab:condition-ground-truth-results}
    \small

    \begin{adjustbox}{center,max width=\textwidth}
    \begin{tabular}{llrrrrr}
        \toprule

        \multicolumn{7}{l}{\textit{Regression coefficients}} \\
        \addlinespace[2pt]
        \multicolumn{2}{l}{Fixed effect}
            & Estimate & SE & & $z$ & $p$ \\
        \midrule
        \multicolumn{2}{l}{Intercept}
            & 1.39 & 0.09 & & 15.51 & $<.001$ \\
        \multicolumn{2}{l}{Condition 1 (Smooth majority)}
            & 0.14 & 0.10 & & 1.41 & .160 \\
        \multicolumn{2}{l}{Condition 2 (Smooth majority)}
            & $-0.07$ & 0.10 & & $-0.66$ & .507 \\
        \multicolumn{2}{l}{Ground truth 1 (Non-smooth galaxy)}
            & $-0.01$ & 0.06 & & $-0.19$ & .851 \\
        \multicolumn{2}{l}{Condition 1 $\times$ Ground truth 1}
            & 0.07 & 0.03 & & 2.14 & .033 \\
        \multicolumn{2}{l}{Condition 2 $\times$ Ground truth 1}
            & 0.64 & 0.04 & & 14.72 & $<.001$ \\

        \addlinespace[6pt]
        \multicolumn{7}{l}{\textit{ANOVA omnibus tests}} \\
        \addlinespace[2pt]
        \multicolumn{2}{l}{Fixed effect}
            & $\chi^2$ & $df$ & & & $p$ \\
        \midrule
        \multicolumn{2}{l}{Intercept}
            & 240.64 & 1 & & & $<.001$ \\
        \multicolumn{2}{l}{Condition}
            & 1.98 & 2 & & & .372 \\
        \multicolumn{2}{l}{Ground truth}
            & 0.04 & 1 & & & .851 \\
        \multicolumn{2}{l}{Condition $\times$ Ground truth}
            & 268.86 & 2 & & & $<.001$ \\

        \addlinespace[6pt]
        \multicolumn{7}{l}{\textit{Post-hoc contrasts}} \\
        \addlinespace[2pt]
        Ground truth & Condition contrast
            & Estimate & SE & $z$ & &  $p$ \\
        \midrule
        Smooth
            & Balanced $-$ Non-smooth-majority
            & $-0.36$ & 0.20 & $-1.83$  & & .402 \\
        Smooth
            & Balanced $-$ Smooth-majority
            & 0.99 & 0.17  & 5.90  & & $<.001$ \\
        Smooth
            & Non-smooth-majority $-$ Smooth-majority
            & 1.35 & 0.19  & 6.98  & & $<.001$ \\
        \addlinespace
        Non-smooth
            & Balanced $-$ Non-smooth-majority
            & 0.77 & 0.17  & 4.61  & & $<.001$ \\
        Non-smooth
            & Balanced $-$ Smooth-majority
            & $-0.57$ & 0.20  & $-2.82$  & & .029 \\
        Non-smooth
            & Non-smooth-majority $-$ Smooth-majority
            & $-1.35$ & 0.20  & $-6.72$  & & $<.001$ \\
        \bottomrule
    \end{tabular}
    \end{adjustbox}

    \vspace{0.5em}
\end{table*}

\begin{table*}[htbp]
    \centering
    \caption{Results of the logistic linear mixed-effects model examining
    trial-level accuracy with no participants excluded from analyses, and post-hoc contrasts.}
    \label{tab:condition-ground-truth-results-second-model}
    \small

    \begin{adjustbox}{center,max width=\textwidth}
    \begin{tabular}{llrrrrr}
        \toprule

        \multicolumn{7}{l}{\textit{Regression coefficients}} \\
        \addlinespace[2pt]
        \multicolumn{2}{l}{Fixed effect}
            & Estimate & SE & & $z$ & $p$ \\
        \midrule
        \multicolumn{2}{l}{Intercept}
            & 1.30 & 0.08 & & 17.05 & $<.001$ \\
        \multicolumn{2}{l}{Condition 1 (Smooth majority)}
            & 0.07 & 0.08 & & 0.90 & .369 \\
        \multicolumn{2}{l}{Condition 2 (Smooth majority)}
            & $-0.04$ & 0.08 & & $-0.56$ & .574 \\
        \multicolumn{2}{l}{Ground truth 1 (Non-smooth galaxy)}
            & 0.02 & 0.06 & & 0.32 & .751 \\
        \multicolumn{2}{l}{Condition 1 $\times$ Ground truth 1}
            & $-0.01$ & 0.03 & & $-0.27$ & .787 \\
        \multicolumn{2}{l}{Condition 2 $\times$ Ground truth 1}
            & 0.58 & 0.03 & & 17.52 & $<.001$ \\

        \addlinespace[6pt]
        \multicolumn{7}{l}{\textit{ANOVA omnibus tests}} \\
        \addlinespace[2pt]
        \multicolumn{2}{l}{Fixed effect}
            & $\chi^2$ & $df$ & & & $p$ \\
        \midrule
        \multicolumn{2}{l}{Intercept}
            & 290.61 & 1 & & & $<.001$ \\
        \multicolumn{2}{l}{Condition}
            & 0.83 & 2 & & & .660 \\
        \multicolumn{2}{l}{Ground truth}
            & 0.10 & 1 & & & .751 \\
        \multicolumn{2}{l}{Condition $\times$ Ground truth}
            & 345.93 & 2 & & & $<.001$ \\

        \addlinespace[6pt]
        \multicolumn{7}{l}{\textit{Post-hoc contrasts}} \\
        \addlinespace[2pt]
        Ground truth & Condition contrast
            & OR & SE & $z$ & & $p$ \\
        \midrule
        Smooth
            & Balanced / Non-smooth-majority
            & 0.62 & 0.10 & $-3.11$ & & .011 \\
        Smooth
            & Balanced / Smooth-majority
            & 1.93 & 0.25  & 5.05 & & $<.001$ \\
        Smooth
            & Non-smooth-majority / Smooth-majority
            & 3.10 & 0.47  & 7.50  & &$<.001$ \\
        \addlinespace
        Non-smooth
            & Balanced / Non-smooth-majority
            & 2.01 & 0.26  & 5.37  & &$<.001$ \\
        Non-smooth
            & Balanced / Smooth-majority
            & 0.62 & 0.10  & $-3.06$  & & .013 \\
        Non-smooth
            & Non-smooth-majority / Smooth-majority
            & 0.31 & 0.05  & $-7.65$  & & $<.001$ \\
        \bottomrule
    \end{tabular}
    \end{adjustbox}

    \vspace{0.5em}
\end{table*}

\begin{table*}[htbp]
    \centering
    \caption{Results of the logistic linear mixed-effects model assessing
    changes in trial-level accuracy across experimental blocks.}
    \label{tab:block-glmm-results}
    \small
    \begin{tabular}{llrrrr}
        \toprule

        \multicolumn{6}{l}{\textit{Regression coefficients}} \\
        \addlinespace[2pt]
        \multicolumn{2}{l}{Fixed effect}
            & Estimate & SE & $z$ & $p$ \\
        \midrule
        \multicolumn{2}{l}{Intercept}
            & 1.13 & 0.10 & 10.89 & $<.001$ \\
        \multicolumn{2}{l}{Condition 1 (Smooth majority)}
            & 0.34 & 0.12 & 2.86 & .004 \\
        \multicolumn{2}{l}{Condition 2 (Smooth majority)}
            & $-0.03$ & 0.12 & $-0.24$ & .808 \\
        \multicolumn{2}{l}{Order block 1 (Third block)}
            & $-0.06$ & 0.08 & $-0.79$ & .430 \\
        \multicolumn{2}{l}{Order block 2 (Third block)}
            & 0.06 & 0.08 & 0.75 & .451 \\
        \multicolumn{2}{l}{Condition 1 $\times$ Order block 1}
            & $-0.13$ & 0.12 & $-1.15$ & .249 \\
        \multicolumn{2}{l}{Condition 2 $\times$ Order block 1}
            & 0.19 & 0.11 & 1.71 & .088 \\
        \multicolumn{2}{l}{Condition 1 $\times$ Order block 2}
            & 0.00 & 0.12 & 0.02 & .984 \\
        \multicolumn{2}{l}{Condition 2 $\times$ Order block 2}
            & $-0.00$ & 0.11 & $-0.04$ & .968 \\

        \addlinespace[6pt]
        \multicolumn{6}{l}{\textit{ANOVA omnibus tests}} \\
        \addlinespace[2pt]
        \multicolumn{2}{l}{Fixed effect}
            & $\chi^2$ & $df$ & & $p$ \\
        \midrule
        \multicolumn{2}{l}{Intercept}
            & 118.50 & 1 & & $<.001$ \\
        \multicolumn{2}{l}{Condition}
            & 9.83 & 2 & & .007 \\
        \multicolumn{2}{l}{Order block}
            & 0.80 & 2 & & .671 \\
        \multicolumn{2}{l}{Condition $\times$ Order block}
            & 3.98 & 4 & & .409 \\
        \bottomrule
    \end{tabular}

    \vspace{0.5em}
    \begin{minipage}{0.95\linewidth}
        \footnotesize
        \textit{Note.} Order block represents each 50-items block.
    \end{minipage}
\end{table*}
\begin{table*}[htbp]
    \centering
    \caption{Results of the logistic linear mixed-effects model examining
    trial-level accuracy as a function of response streak length and
    experimental condition.}
    \label{tab:response-streak-results}
    \small

    \begin{adjustbox}{center,max width=\textwidth}
    \begin{tabular}{llrrrrr}
        \toprule

        \multicolumn{7}{l}{\textit{Regression coefficients}} \\
        \addlinespace[2pt]
        \multicolumn{2}{l}{Fixed effect}
            & Estimate & SE & & $z$ & $p$ \\
        \midrule
        \multicolumn{2}{l}{Intercept}
            & 0.85 & 0.07 & & 11.76 & $<.001$ \\
        \multicolumn{2}{l}{Response streak length}
            & $-0.03$ & 0.01 & & $-3.23$ & .001 \\
        \multicolumn{2}{l}{Condition 1 (Smooth majority)}
            & 0.61 & 0.07 & & 9.33 & $<.001$ \\
        \multicolumn{2}{l}{Condition 2 (Smooth majority)}
            & $-0.28$ & 0.07 & & $-4.20$ & $<.001$ \\
        \multicolumn{2}{l}{Response streak length $\times$ Condition 1}
            & 0.06 & 0.02 & & 3.89 & $<.001$ \\
        \multicolumn{2}{l}{Response streak length $\times$ Condition 2}
            & $-0.02$ & 0.01 & & $-1.72$ & .086 \\

        \addlinespace[6pt]
        \multicolumn{7}{l}{\textit{ANOVA omnibus tests}} \\
        \addlinespace[2pt]
        \multicolumn{2}{l}{Fixed effect}
            & $\chi^2$ & $df$ & & & $p$ \\
        \midrule
        \multicolumn{2}{l}{Intercept}
            & 138.21 & 1 & & & $<.001$ \\
        \multicolumn{2}{l}{Response streak length}
            & 10.40 & 1 & & & .001 \\
        \multicolumn{2}{l}{Condition}
            & 87.02 & 2 & & & $<.001$ \\
        \multicolumn{2}{l}{Response streak length $\times$ Condition}
            & 15.62 & 2 & & & $<.001$ \\

        \addlinespace[6pt]
        \multicolumn{7}{l}{\textit{Post-hoc contrasts of trends}} \\
        \addlinespace[2pt]
        \multicolumn{2}{l}{Condition contrast}
            & Estimate & SE & $df$ & $z$ & $p$ \\
        \midrule
        \multicolumn{2}{l}{B $-$ N}
            & 0.08 & 0.03 & $\infty$ & 3.31 & .003 \\
        \multicolumn{2}{l}{B $-$ S}
            & 0.11 & 0.03 & $\infty$ & 3.89 & $<.001$ \\
        \multicolumn{2}{l}{N $-$ S}
            & 0.02 & 0.02 & $\infty$ & 1.10 & .818 \\
        \bottomrule
    \end{tabular}
    \end{adjustbox}

    \vspace{0.5em}
    \begin{minipage}{0.95\linewidth}
        \footnotesize
        \textit{Note.} Response streak length indicates the number of
        consecutive identical responses up to the current trial.
    \end{minipage}
\end{table*}

\end{document}